\documentclass[aps,pra,showpacs,twocolumn,amsmath,amssymb]{revtex4}

\usepackage[english]{babel}
\usepackage{latexsym}
\usepackage{graphics,psfrag}
\usepackage{epsfig}
\usepackage{color}

\def\be{\begin{equation}}
\def\ee{\end{equation}}
\def\bea{\begin{eqnarray}}
\def\eea{\end{eqnarray}}
\def\bi{\begin{itemize}}
\def\ei{\end{itemize}}
\def\bin{\begin{enumerate}}
\def\ein{\end{enumerate}}
\def\la{\langle}
\def\ra{\rangle}

\newcommand{\vect}[1]{\mathbf{#1}}
\newcommand{\rmd}{\mathrm{d}}

\begin{document}
\title{Self-localization of a small number of Bose particles 
in a superfluid Fermi system}

\author{Katarzyna Targo\'nska}
\affiliation{
Instytut Fizyki imienia Mariana Smoluchowskiego and
Mark Kac Complex Systems Research Center, 
Uniwersytet Jagiello\'nski, ulica Reymonta 4, PL-30-059 Krak\'ow, Poland}

\author{Krzysztof Sacha}
\affiliation{
Instytut Fizyki imienia Mariana Smoluchowskiego and
Mark Kac Complex Systems Research Center, 
Uniwersytet Jagiello\'nski, ulica Reymonta 4, PL-30-059 Krak\'ow, Poland}

\date{\today}

\begin{abstract}
We consider self-localization of a small number of Bose particles 
immersed in a large homogeneous superfluid mixture of fermions in three 
and one dimensional spaces. Bosons distort the density of surrounding fermions 
and create a potential well where they can form a bound state 
analogous to a small polaron state. 
In the three dimensional volume we observe the self-localization for
repulsive interactions between bosons and fermions. 
In the one dimensional case bosons self-localize as well as for attractive 
interactions forming, together with a pair of fermions at the bottom of the 
Fermi sea, a vector soliton. 
We analyze also thermal effects and show that small 
non-zero temperature affects 
the pairing function of the Fermi-subsystem and has little influence on the
self-localization phenomena.
\end{abstract}


\maketitle

\section{Introduction}

Ultra-cold atomic gases offer possibilities for
realizations of complex mathematical models used in different
fields of physics with an unprecedented level of the experimental control
\cite{Lewenstein2007,rmp08}.
For example, condensed matter phenomena like the superfluid-Mott insulator
transition and the Bose-glass phase or the Anderson localization effects can be
experimentally investigated \cite{bloch02,Fallani2007,Billy2008,ingu2008}. 
Fermionic gases, in particular Fermi 
superfluids, have received a lot of attention, especially after the observation of 
the transition between 
the superfluid Bardeen-Cooper-Schrieffer (BCS) pairs and the Bose-Einstein 
condensate (BEC) of diatomic molecules
\cite{regal2004,varonna2006}.

The behavior of a small {\it object} immersed in degenerate quantum gases has
been investigated by several authors 
\cite{Kalas2006,Cucchietti2006,Sacha2006,Bruderer2008a,Bruderer2008b,
Sanamore2008,Roberts2009,Novikov2009,Tempere2009,Novikov2010}. 
For example, weak interactions between 
a single impurity atom and particles of a large BEC can be described by the
perturbation theory. For stronger  
interactions an effective mass of an impurity atom diverges indicating the breakdown of
the perturbation approach and the self-localization of the impurity {\it object} 
in a close analogy to the small polaron problem, i.e. localization of an
electron in a surrounding cloud of lattice distortions \cite{Mahan}. 
In ultra-cold fermionic
gases an example of polaron effects with a small number of spin-up fermions 
immersed in a large cloud of spin-down Fermi particles has been studied 
theoretically
\cite{Chevy2006,Combescot2007,Punk2007,Prokofev2008a,Prokofev2008b,Punk2009} 
and recently realized experimentally \cite{Schirotzek2009,Nascimbene2009}. 
Employing a Feshbach resonance, that allows tuning the
interaction strength between atoms, experimentalists have been able to
investigate a transition from the nearly non-interacting case, through 
the polaron regime to the limit where pairs of unlike fermions form tightly 
bound molecules.

In the present publication we consider a small number of Bose particles
immersed in a large, homogeneous, superfluid and balanced mixture of spin-up 
and
spin-down fermions and analyze the
self-localization phenomenon. Another limit, investigated already in the
literature, 
concerns Bose-Fermi mixtures with a number of bosons comparable to (or even 
larger than) a number of fermions and effects of the phase separation
\cite{Molmer1998,Viverit2000,Yip2001,Roth2002,Pu2002,Adhikari2007,Bhongale2008,
Luhmann2008,Lee,Ramachandhram,Mashayekhi}. 
The latter
corresponds to instability of a homogeneous solution when boson-fermion 
interaction reaches a critical strength. In the case of small boson numbers, 
the boson-boson 
interactions can be neglected and the uniform density solution is unstable as
soon as the boson-fermion coupling constant becomes non-zero. 
However, this does not mean the self-localization of Bose particles. 
We show that 
the self-localization 
takes place for stronger interactions when the boson-fermion coupling constant 
is greater than a non-zero critical value. 

The possibility of solitonic behavior in 
Bose-Fermi mixtures with fermions both in the normal and superfluid states has
been investigated in the literature
\cite{Karpiuk2004,Karpiuk2006,Adhikari2005,Adhikari2007}. 
For a large number of bosons, if the 
attractive boson-fermion interaction is sufficiently strong, the boson-boson
repulsion may be outweighed and the whole Bose and Fermi clouds reveal solitonic
behavior. We consider Bose-Fermi mixtures in the opposite limit of
small boson numbers. In that regime different kind of solitons exists. 
Indeed, in the 1D case description of the system may be reduced to a simple 
model where bosons and a single pair of fermions at the bottom of the Fermi sea 
are described by a vector soliton solution.

The paper is organized as follows. In Sec.~\ref{model} we introduce the model
used in the description of Bose-Fermi mixtures. The results for the case of  
three-dimensional (3D) and 1D spaces are collected 
in Sec.~\ref{results} and we conclude in Sec.~\ref{conclusions}.

\section{Model description}
\label{model}

Let us consider a small number $N_b$ of 
bosonic atoms in the Bose-Einstein condensate 
state immersed in a homogeneous, 
dilute and balanced mixture of fermions in two different internal 
spin states 
in a 3D volume. 
Interactions of ultra-cold atoms can be described via contact potentials 
${\cal V}_{ij}(\vect r)=g_{ij}\delta(\vect r)$
with strengths given 
in terms of $s$-wave scattering lengths $a_{ij}$ as 
$g_{ij}=\frac{2\pi \hbar^2 a_{ij}}{m_{ij}}$, 
where $m_{ij}$ stands for a reduce mass of a pair of interacting atoms. 
In our model we consider attractive interactions between 
fermions in different spin states, i.e. negative coupling constant $g_{ff}$. 
Interactions between bosons and fermions are 
determined by the spin-independent parameter $g_{bf}$. 
We neglect mutual interactions 
of bosonic atoms in the assumption that either their density 
remains sufficiently small or the coupling constant is negligible.

The system is described by the following Hamiltonian
\bea \label{H}
 \hat{H} &=& \int  \rmd^3 r \left[ \hat{\Psi } _{b} ^{\dagger}
 \left(-\frac{\hbar^2}{2m_{b}}\nabla^2 \right) \hat {\Psi}_{b}+ 
 \displaystyle \sum_{s=+,-} \left( {\hat{\Psi}_{f,s}^{\dagger} 
 H_{0} \hat {\Psi}_{f,s}} \right. \right. \cr
 && \left. \left. - \frac {|g_{ff}|}{ 2} {\hat{\Psi}_{f,s}^{ \dagger} 
 \hat{\Psi}_{f,-s}^{\dagger} \hat{\Psi}_{f,-s} \hat{\Psi}_{f,s} }
  +g_{bf} {\hat{\Psi}_{f,s}^{\dagger} \hat{\Psi}_{f,s}}{\hat{\Psi}_{b}^{\dagger} 
  \hat{\Psi}_{b}} \right) \right], 
  \cr  &&
 \eea
where $H_{0}=-\frac{\hbar^2}{2m_{f}}\nabla^{2}-\mu$.
$\hat{\Psi } _{b}$ and $\hat{\Psi}_{f,s}$  
refer, respectively, to the field operators of bosonic and fermionic atoms 
where $s\in\{+,-\}$ indicates a spin state. $\mu$ 
stands for the chemical potential of the Fermi sub-system 
and $m_b$ and $m_f$ are masses of bosons and
fermions, respectively.
 
We look for a thermal equilibrium state assuming that the Bose
and Fermi sub-systems are separable. For instance in the limit of zero
temperature it is given by a product ground state 
\be
\label{produkt}
|{\Psi}\ra= |\psi\ra_{f} |\phi \ra_{b}.
\ee
We also postulate that the Fermi sub-system can be described by the BCS mean-field 
approximation \cite{varonna2006} with the paring field 
$\Delta(\vect{r})=|g_{ff}|\left<\hat{\psi}_{f,+}
\hat{\psi}_{f,-}\right>$ 
and the Hartree-Fock potential 
$W(\vect{r}) = -|g_{ff}|\left<\hat{\psi}_{f,+}^{\dagger}\hat{\psi}_{f,+}\right>
=-|g_{ff}|\left<\hat{\psi}_{f,-}^{\dagger}\hat{\psi}_{f,-}\right>$ 
affected by a potential proportional to the density of bosons 
$N_b|\phi(\vect r)|^{2}$.
Assuming a spherical symmetry of particle densities, 
the description of the system reduces to 
the Bogoliubov-de Gennes equations for fermions
\bea\label{bg}
\left({H_{0}+W +g_{bf}N_b|\phi|^{2}}\right)u_{nlm}+\Delta v_{nlm}&=&E_{nl} u_{nlm}\cr
\Delta^* u_{nlm}-\left({H_{0}+W +g_{bf}N_b|\phi|^{2}} \right
)v_{nlm}&=&E_{nl}v_{nlm}, \cr &&
\eea
where $l$ and $m$ stand for angular momentum quantum numbers and
\bea
W &=&-|g_{ff}|\displaystyle \sum_{nlm}\left[f_{nl}|u_{nlm}(\vect r)|^2  
+(1-f_{nl})|v_{nlm}(\vect r)|^{2}\right], 
\cr && \\
 \Delta &=& |g_{ff}| \displaystyle \sum_{nlm} (1-2f_{nl}) u_{nlm}
 (\vect{r}) v^{*}_{nlm} (\vect{r}), 
\label{delta} 
\eea 
with the Fermi-Dirac distribution
\be
f_{nl}=\frac{1}{\exp(E_{nl}/k_BT)+1},
\ee
which have to be solved
together with the Gross-Pitaevskii equation for bosons
\be\label{bosony}
{\left[		-\frac{\hbar^2}{2m_b}\nabla^{2}+V(\vect r)\right]\phi(\vect r)}=
{\mu_{b}\phi(\vect r)},
\ee
where 
\be\label{veff}
V(\vect r)=-\frac{2g_{bf}}{|g_{ff}|}W(\vect r)=g_{bf}\rho_{f}(\vect r).
\ee
The effective potential $V(\vect r)$ for bosons comes from contact interactions between 
bosons and fermions. $\rho_f$ is density of fermions 
and $\mu_{b}$ is the chemical potential for bosons. We consider the 
temperature much lower than the critical temperature for Bose-Einstein 
condensation therefore we can neglect thermal excitations of bosons. 

The coupled equations (\ref{bg}) and (\ref{bosony}) are solved numerically
in a self-consisted manner. 
In the calculations we adopt 
\bea\label{units}
E_0&=&2E_F=\frac{\hbar^2k_F^2}{m_f}, \cr
l_0&=&\frac{1}{k_F},
\eea 
units for energy and length, respectively, where 
$k_F=(3\pi^2 n_0 )^{1/3}$ is the Fermi wave-number of a uniform ideal Fermi
gas of density $n_0$. In these units the coupling constants are the following
\bea
g_{ff}&=&4\pi \;k_Fa_{ff}, \cr
g_{bf}&=&2\pi \;k_Fa_{bf}\left(1+\frac{m_f}{m_b}\right),
\eea
and we deal with six independent parameters in the system: number of bosons $N_b$,
chemical potential of Fermi sub-system $\mu$, ratio of the masses  
$\frac{m_b}{m_f}$, scattering lengths $k_Fa_{ff}$ and $k_Fa_{bf}$ and
radius $R$ of the 3D volume we consider.

In the 3D case the coupling constant $g_{ff}$ in $\Delta$
[Eq.~(\ref{delta})] has to be regularized in order to avoid ultraviolet
divergences. That is, $g_{ff}\rightarrow g_{\rm eff}$ where
\begin{equation}
\label{regular}
\frac{1}{|g_{\rm
eff}|}=\frac{1}{|g_{ff}|}-\frac{1}{2\pi^2}\left(
\frac12\ln\frac{\sqrt{E_C}+\sqrt{\mu}}{
\sqrt{E_C}-\sqrt{\mu}}-
\sqrt{\frac{E_C}{\mu}}\right).
\end{equation}
The logarithmic term in (\ref{regular}) results from the sum over
Bogoliubov modes corresponding to the energy above a numerical cutoff $E_C$
performed in the spirit of the local density approximation, see
\cite{Bruun1999,Bulgac2002,Grasso2003,Niederberger2009} for details. 

\section{Results}
\label{results}

Without interactions between bosons and fermions the ground state of the
system corresponds to uniform particle densities. 
For the non-zero coupling constant $g_{bf}$,
the uniform solution become unstable and, depending on the sign of $g_{bf}$,
the bosonic and fermionic clouds tend to separate from each other 
or try to stick together. For sufficiently strong interactions, the effect of 
the self-localization may be expected (see the similar problem in the case of 
an impurity atom immersed in a large Bose-Einstein condensate considered in 
Ref.~\cite{Kalas2006,Cucchietti2006,Sacha2006}). 
Indeed, for $g_{bf}>0$ bosons repel 
fermions and create a potential well in their vicinity where they may localize
if the well is sufficiently large. For attractive interactions the
density of fermions increases in the vicinity of Bose particles. 
Due to the fact that $g_{bf}<0$, the bosons experience the density deformation 
in a form of a potential well and they may localize.

We begin with the 3D model and focus on the repulsive boson-fermion 
interactions. Analysis of both zero-temperature limit and thermal effects are
performed. Then we consider the 1D case where 
the self-localization phenomenon may be related to
the presence of a vector soliton solution.

\subsection{Three dimensional model}

\begin{figure}
\centering
\includegraphics*[width=0.9\linewidth]{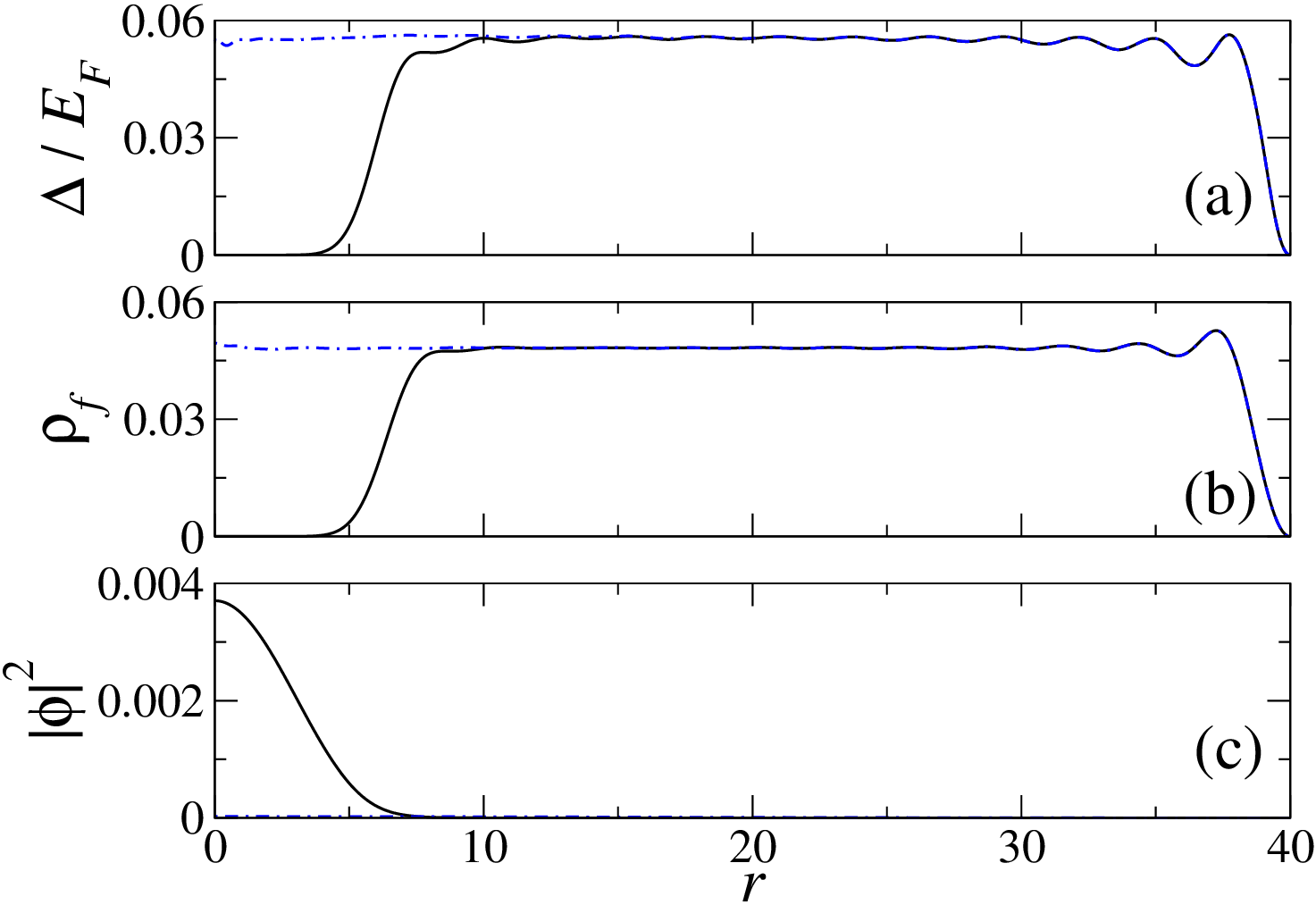}
\caption{(Color online) Self-localization of $^{23}$Na atoms in 
a superfluid mixture of $^{40}$K atoms. Panel (a) shows the pairing function
$\Delta(r)$, panel (b) fermion density $\rho_f(r)$ and panel (c) density of 
bosons $|\phi(r)|^2$. Solid black lines
correspond to boson-fermion interaction strength 
$g_{bf}=10$ and dotted-dashed blue lines to $g_{bf}=0$. 
In panel (c) the dotted-dashed blue line is 
hardly visible because for $g_{bf}=0$ bosons are delocalized and their 
density very small.
Number of bosons $N_b=100$ and fermions $N_f\approx12000$ (chemical potential 
$\mu=E_F$) and fermion-fermion coupling constant $g_{ff}=-5.5$.
}
\label{one}
\end{figure}
\begin{figure}
\centering
\includegraphics*[width=0.9\linewidth]{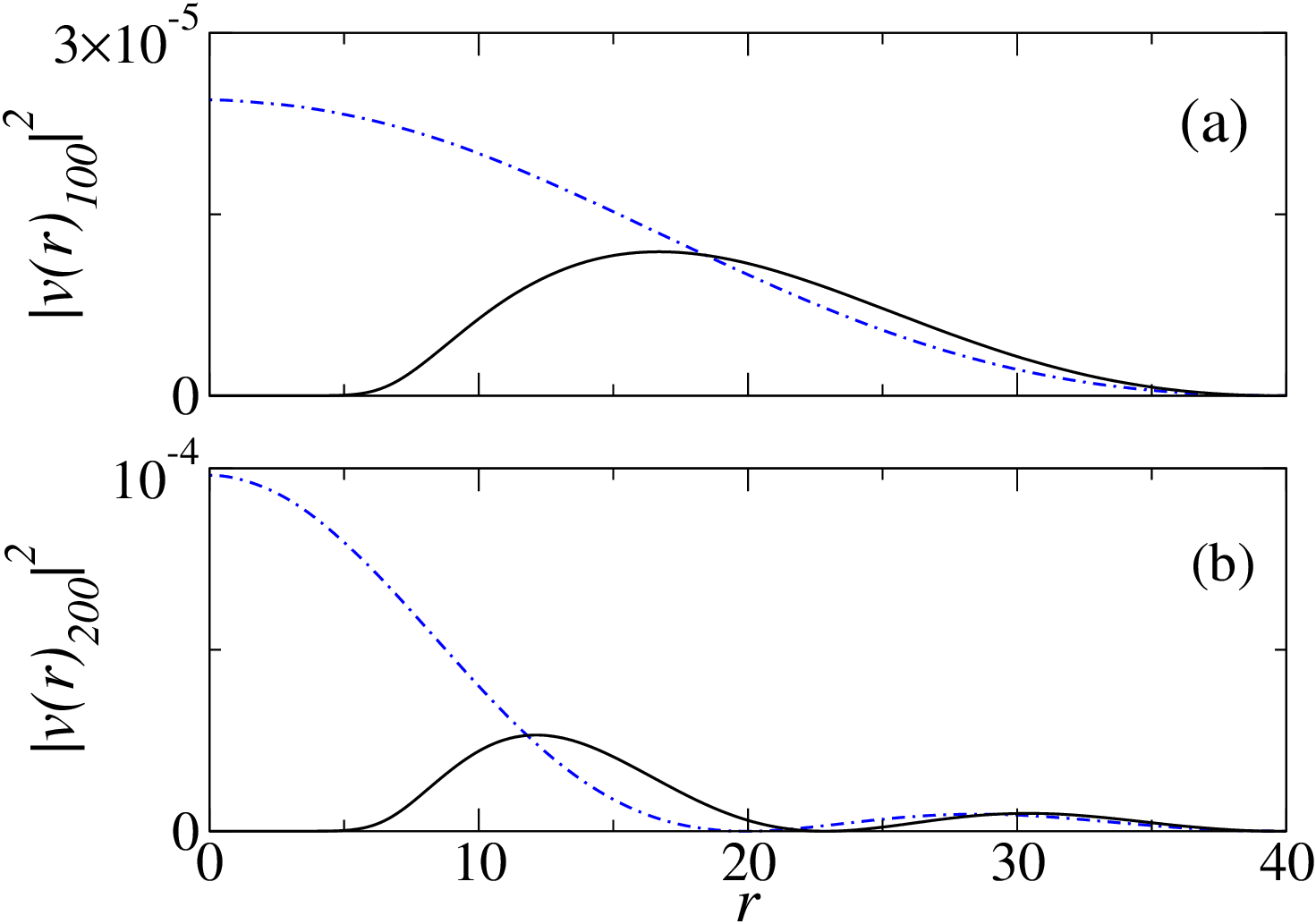}
\caption{(Color online) Probability densities $|v_{nlm}(r)|^2$ 
of two fermion pairs at the bottom of the Fermi sea with angular 
momentum $l=0$. Panel (a) corresponds to the ground state ($n=1$) of the 
radial degree of freedom and panel (b) to the first excited state ($n=2$). 
Solid black lines correspond to boson-fermion interaction strength 
$g_{bf}=10$ and dotted-dashed blue lines to $g_{bf}=0$. 
All parameters are the same as in Fig.~\ref{one}. 
}
\label{two}
\end{figure}
\begin{figure}
\centering
\includegraphics*[width=0.9\linewidth]{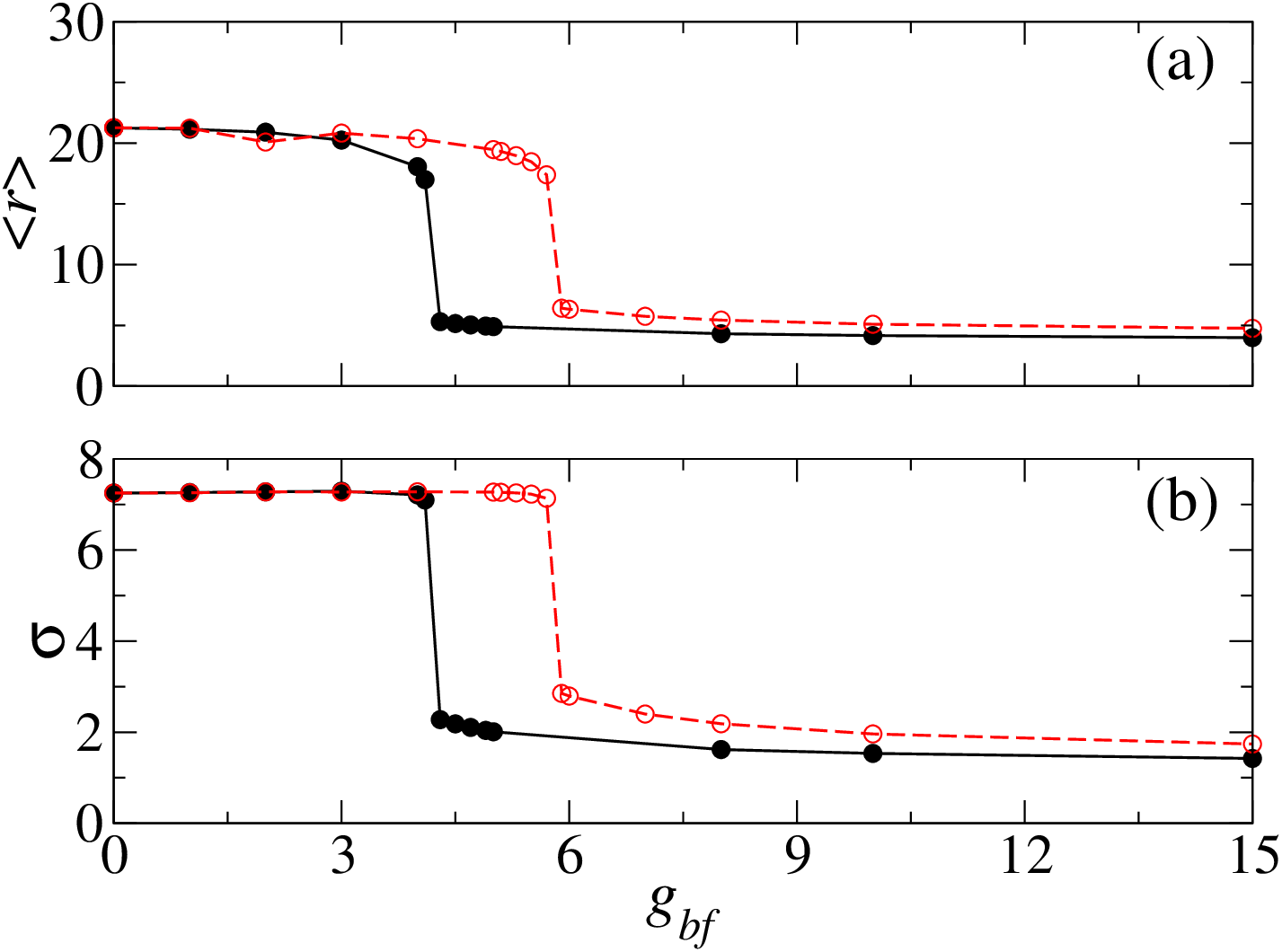}
\caption{(Color online) The average radius of the Bose cloud $\la r\ra$ [panel (a)]
and the standard deviation $\sigma=(\la r^2\ra-\la r\ra^2)^{1/2}$ [panel (b)] versus
boson-fermion coupling constant $g_{bf}$. Black full symbols correspond to a
mixture of $^{23}$Na and $^{40}$K atoms while red open symbols to 
a mixture of $^{7}$Li and $^{40}$K atoms. Note the abrupt transitions to localized
states when critical values of $g_{bf}$ are reached.
All the other parameters are the same as in Figs.~\ref{one}-\ref{two}. 
}
\label{three}
\end{figure}
\begin{figure}
\centering
\includegraphics*[width=0.9\linewidth]{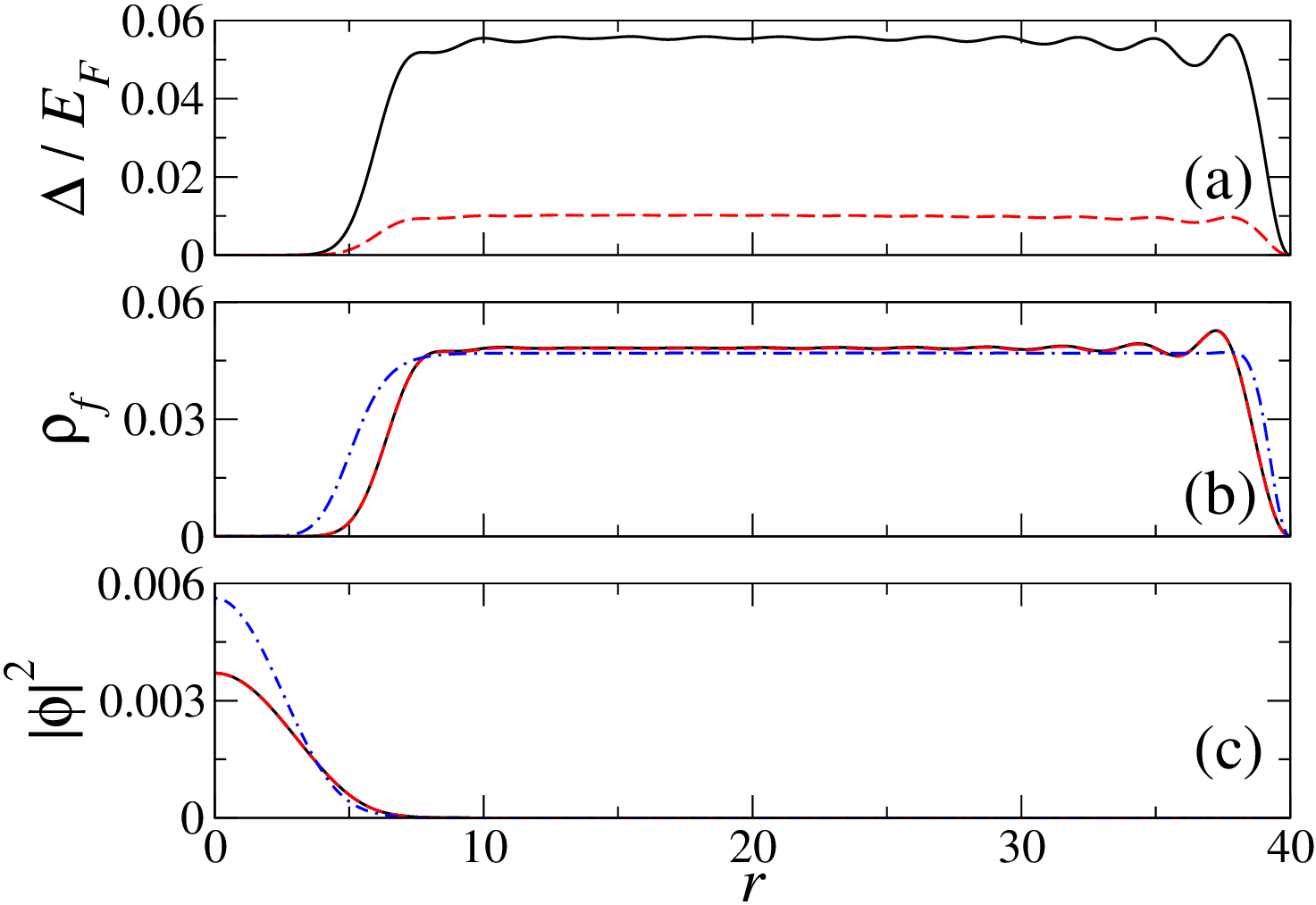}
\caption{(Color online) 
 Self-localization of $^{23}$Na atoms in 
a superfluid mixture of $^{40}$K atoms for non-zero temperature. 
Panel (a) shows the pairing function $\Delta(r)$, panel (b) fermion density 
$\rho_f(r)$ and panel (c) density of bosons $|\phi(r)|^2$. 
Solid black lines correspond to $T=0$ and $\mu=E_F$, red dashed lines to 
$T=0.028 T_F$ and $\mu =E_F $, blue dotted-dashed lines to $T=6T_F$ and $\mu=0.16 E_F$.
Boson-fermion interaction strength $g_{bf}=10$, fermion-fermion coupling 
constant $g_{ff}=-5.5$ and number of bosons $N_b=100$ and fermions $N_f\approx 12000$.
In panel (a) the dotted-dashed blue line is not visible because for 
$T=6 T_F$ the pairing function is equal zero. 
In panels (c) and (d) the solid 
black and dashed red lines are hardly distinguishable.
}
\label{four}
\end{figure}

Figure~\ref{one} shows the  densities of bosons and fermions and the pairing function
corresponding to the ground state of the system for
$g_{bf}=0$ and $g_{bf}=10$. Without boson-fermion interactions the
quantities are flat and uniform (except a small region close to the edge of the
3D volume due to assumed open boundary conditions).
However, when the considerable interactions are turned on it becomes energetically favorable
to separate bosons and fermions, the $\rho_f(\vect r)$ is 
depleted around the
center and bosons form a bound state localized in small area around $\vect r=0$.
It is clear, that the localization effect is the result of boson-fermion
interactions. It relies on a local deformation of the density of fermions 
and is not affected by the boundary conditions.

The response of the Fermi sub-system to bosons, that tend to localize, can be
investigated by monitoring deformation of the Bogoliubov quasi-particle modes.
The density of fermions is the sum of the Bogoliubov modes $\rho_f(\vect r)=2\sum_{nlm}|v_{nlm}(\vect r)|^2$.
The modes with zero angular momentum contribute only to the density around 
$\vect r=0$. Consequently, the modification of these modes is primarily responsible 
for preparation of the potential well in which bosons localize. In
Fig.~\ref{two} we illustrate the deformation of two modes with $l=0$ 
corresponding to 
fermions at the bottom of the Fermi sea but we should keep in mind
that all modes with $l=0$ 
become affected by the interactions with bosons. The deformation of modes for
fermions at the Fermi level is reflected by a change of a shape of the pairing
field visible in Fig.~\ref{one}, 
because those modes contribute mainly to $\Delta(\vect r)$.

The interaction of fermions and the {\it impurity} Bose particles influences 
the pairing function $\Delta$ only locally,
see Fig.~\ref{one}. It implies that the superfluidity is not destroyed
even when the interaction is so strong that the localization 
of the {\it impurity} object takes place.

The data in Figs.~\ref{one}-\ref{two} 
are related to $N_b=100$ $^{23}$Na atoms and the mixture of $N_f\approx 12000$
$^{40}$K atoms (chemical potential $\mu=E_F$) in two different hyperfine 
states. We set the scattering lengths $g_{ff}=-5.5$ and $g_{bf}=10$ with
the assumption that they can be realized by the use of the Feshbach
resonances (e.g. magnetic resonance for fermions and optical resonance 
between bosons and fermions \cite{varonna2006,theis2004}). 
In Fig.~\ref{three} we show the average radius 
of the Bose cloud $\la r\ra$ and the standard deviation 
$\sigma=(\la r^2\ra-\la r\ra^2)^{1/2}$ as a function of the coupling 
constant $g_{bf}$. The self-localization means that both $\la r\ra$ and $\sigma$ 
are much smaller than the radius of the 3D volume. One can see that
there is a critical non-zero value of $g_{bf}$ leading to the
self-localization. This critical $g_{bf}$ 
is different from the critical value for the instability of 
the homogeneous solution (i.e. phase separation condition). The latter, 
for the case without boson-boson interactions, corresponds to $g_{bf}>0$. If we
replace the sodium atoms by $^{7}$Li atoms, it turns out that the critical value of
$g_{bf}$ for the self-localization increases. This is, because compressing
the cloud of light lithium particles costs more energy than in the case of heavier 
sodium atoms. 

A small non-zero temperature mostly affects superfluidity and has little effect on
the self-localization phenomenon. Indeed, in Fig.~\ref{four} we see that even for 
$T=0.028 T_F$ when the pairing function is very small 
the densities of bosons and fermions hardly change. 
Increasing temperature to $T=T_F$ (which is still much smaller than
critical temperature for Bose-Einstein condensation of $N_b=100$ bosonic atoms
localized in a volume of the radius $\la r\ra \approx 4$, i.e.  $T_{\rm BEC}\approx 6T_F$)
we observe effects of thermal fluctuations in the fermion density and 
a modification of the density of bosons but the self-localization persists.
Thus, bosons self-localize both for the normal and superfluid phase of 
the Fermi sub-system. 

We have considered the repulsive boson-fermion interaction. For the attractive
interaction we do not observe the self-localization regardless on the phase of
the Fermi sub-system. For $g_{bf}<0$ the particle densities may 
collapse to Dirac-delta distributions. For sufficiently small $|g_{bf}|$
a metastable state may appear. However, it turns out that 
the existence of such a 
metastable state is not the result of self-localization in the system. 
Indeed, it is an effect of a compromise between the requirement of minimal 
kinetic energies and restrictions related to the boundary conditions.
In the following we consider a 1D model where there is no problem with 
the collapse of the densities and show that Bose particles can localize 
in the Fermi sub-system for attractive boson-fermion interactions too.

\subsection{One dimensional model}

\begin{figure}
\centering
\includegraphics*[width=0.9\linewidth]{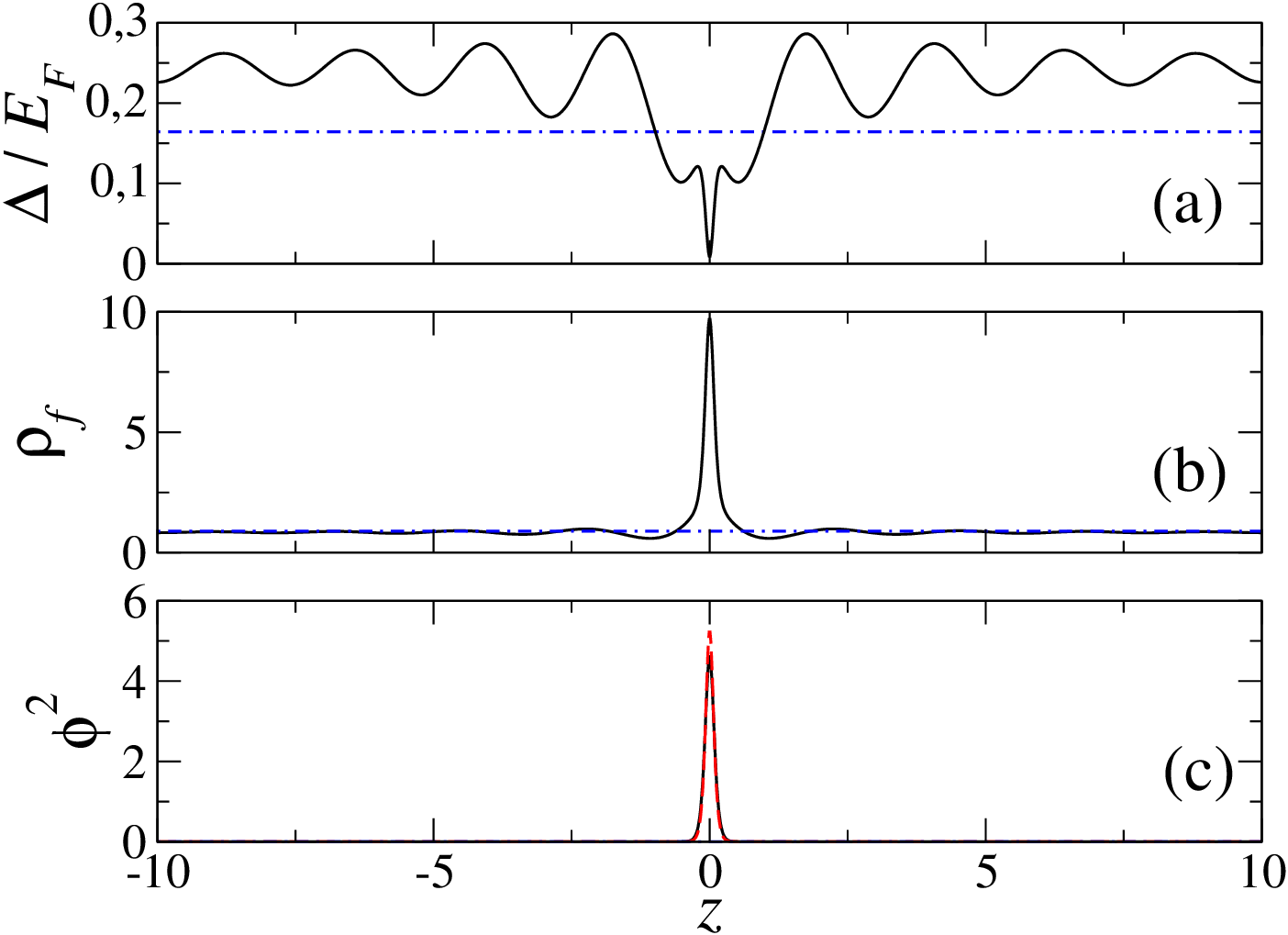}
\caption{(Color online) 
Self-localization of a single boson ($N_b=1$) in a superfluid mixture 
of fermions in 1D space. Panel (a) shows the pairing function $\Delta(z)$, 
panel (b) 
fermion density $\rho_f(z)$ and panel (c) boson density $|\phi(z)|^2$. 
Solid black lines correspond to boson-fermion interaction strength 
$g_{bf}=-20$ and dotted-dashed blue lines to $g_{bf}=0$. 
Number of fermions $N_f\approx 20$ (chemical potential $\mu=E_F$) and 
fermion-fermion coupling constant $g_{ff}=-1$. Ratio of masses of Bose and 
Fermi particles $\frac{m_b}{m_f}$ fulfills Eq.~(\ref{msol}). 
The configuration space extends from $z=-10$ to $z=10$.
In panel (c) the dotted-dashed blue line is 
hardly visible, because the boson is delocalized and its density 
very small for $g_{bf}=0$. Red dashed line in panel (c) indicates the solitonic solution
Eq.~(\ref{soliton}).
}
\label{five}
\end{figure}
\begin{figure}
\centering
\includegraphics*[width=0.9\linewidth]{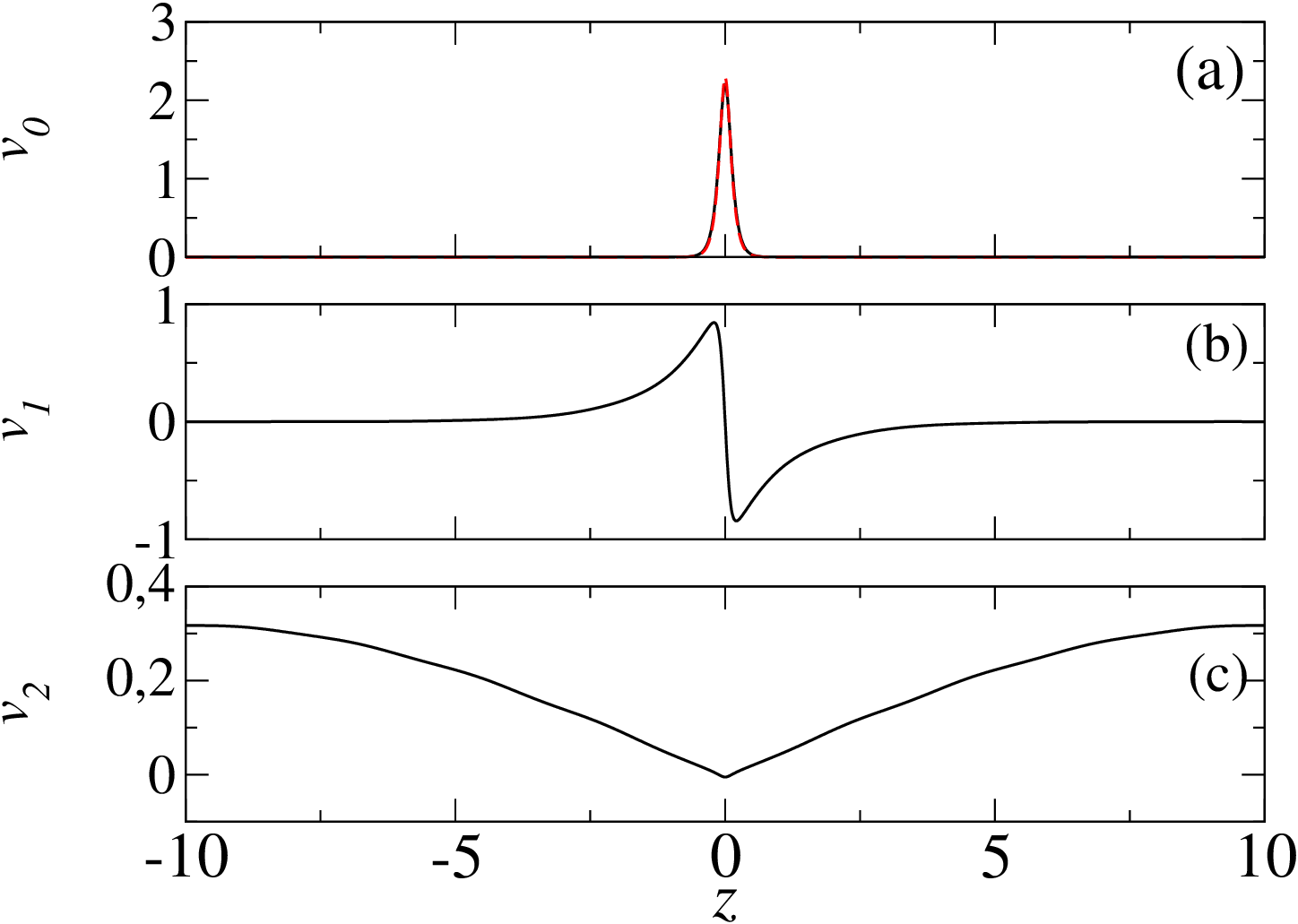}
\caption{(Color online)   
Bogoliubov modes $v_{k}(z)$ corresponding to
fermion pairs located close to the bottom of the Fermi sea.
Panel (a) is related to the pair of fermions at the bottom of the Fermi sea, 
panel (b) and (c) to the next pairs.
Solid black lines correspond to the numerical solutions. 
Red dashed line in panel (a) indicates solitonic solution Eq.~(\ref{soliton}).
All the others parameters are the same as in Fig.~\ref{five}. 
}
\label{six}
\end{figure}
\begin{figure}
\centering
\includegraphics*[width=0.9\linewidth]{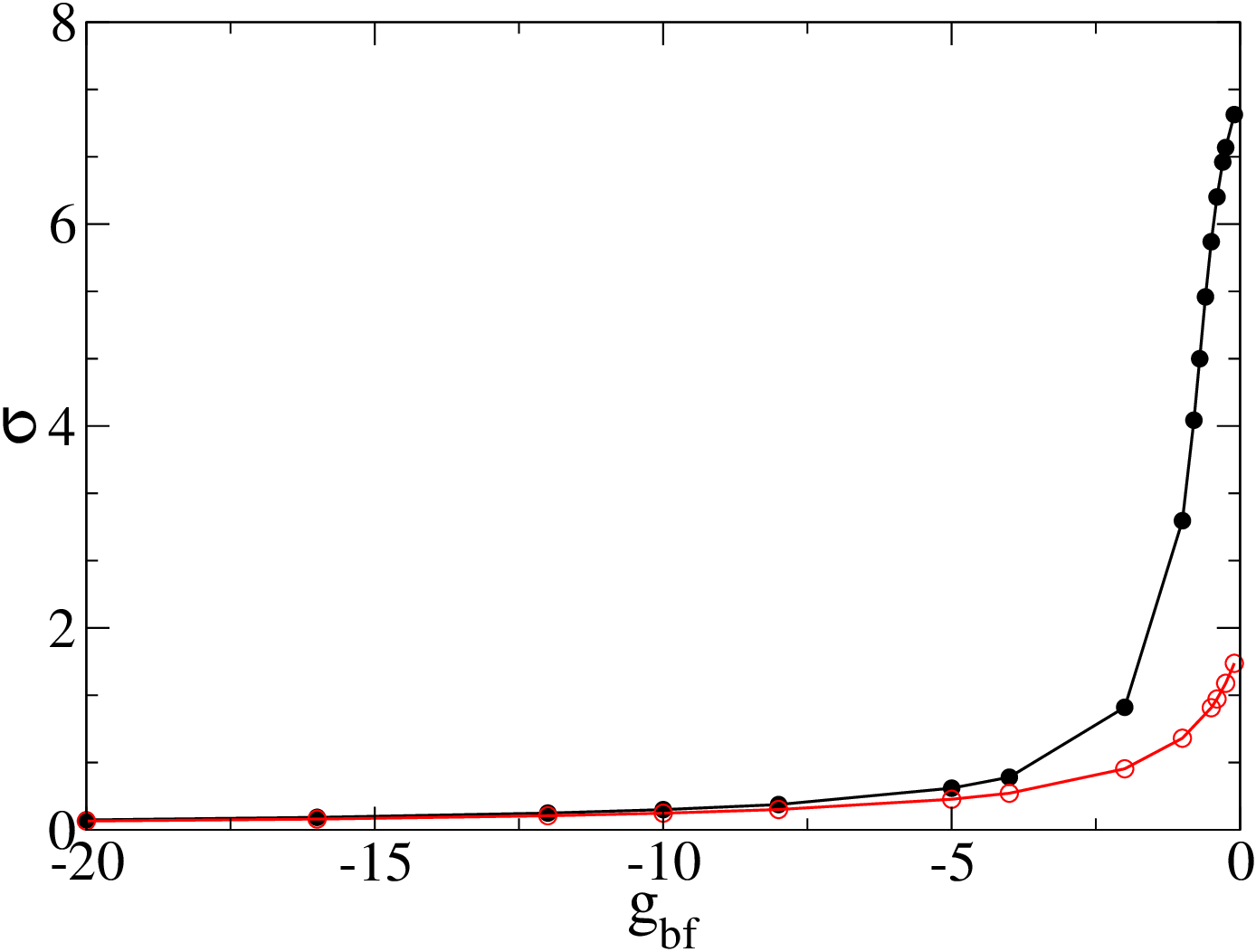}
\caption{(Color online) Width of the boson density, i.e. 
$\sigma=(\la z^2\ra-\la z\ra^2)^{1/2}$, versus boson-fermion coupling constant 
$g_{bf}$. 
Black full symbols correspond to the numerical values and red open symbols to the 
solutions Eq.~(\ref{soliton}). The configuration space extends from
$z=-20$ to $z=20$. All the others parameters are the same as in Fig.~\ref{five}. }
\label{seven}
\end{figure}

If in $x$ and $y$ directions we apply harmonic potentials of 
frequency $\omega_\perp$ and $\hbar\omega_\perp$ is much greater 
than the chemical potentials, the 3D system becomes effectively 
one-dimensional. Then, the description reduces
to the 1D version of Eqs.~(\ref{bg})-(\ref{veff}) with the following
coupling constants [in the units (\ref{units})]
\bea\label{1dcoup}
g_{ff}^{\rm 1D}&=&\frac{2m_f\omega_\perp}{\hbar k_F}a_{ff}, \cr
g_{bf}^{\rm 1D}&=&\frac{2m_f\omega_\perp}{\hbar k_F}a_{bf},
\eea
which have been obtained assuming that the $x$ and $y$ degrees of freedom of 
each atom are in the ground states of the harmonic potentials. In the 1D case 
there is no ultraviolet divergence and the pairing function does not require
regularization. Nevertheless numerical simulations converge much faster if the 
Bogoliubov modes, above a numerical cut-off energy $E_c$, are included in the
spirit of the local density approximation. That is, the coupling constant in
$\Delta$ is substituted by
\be 
\frac{1}{|g_{\rm eff}^{\rm 1D}|}=\frac{1}{|g_{ff}^{\rm 1D}|}-
\frac{1}{2\pi}\ln\frac{\sqrt{E_C}+\sqrt{\mu}}{
\sqrt{E_C}-\sqrt{\mu}}.
\ee

For repulsive boson-fermion interactions, we observe the self-localization 
of bosons with the behaviour of the particle densities
similar as in the 3D case. Therefore we focus on attractive
interactions only. 
Figure~\ref{five} shows the results for $g_{bf}=-20$, obtained with periodic boundary conditions for fermions and open boundary conditions for bosons.
For the attractive interactions bosons and
fermions try to stick together which leads 
to an increase of the fermion density in the
vicinity of the boson concentration 
and the creation of a potential well for 
localization of Bose particles. 

Analyzing the Bogoliubov modes $v_k(z)$
(see Fig.~\ref{six}) we find out that the probability density $v_0^2(z)$ of
a pair of fermions at the bottom of the Fermi sea becomes strongly 
localized. The Bogoliubov mode $v_1(z)$ of the next fermion 
pair forms also a bound state. Since 
$v_1(z)$ is an antisymmetric function it 
is nearly zero in the area around $z=0$.
Probability densities of other fermions are deformed and almost all of 
them drop to zero in the region where $v_0(z)$ is localized. 
This may be interpreted as a realization of 
the Pauli exclusion rule. In the BCS limit only
particles close to the Fermi level contribute to the pairing function $\Delta$ 
and there is practically no contributions from fermions located deeply in the
Fermi sea. Therefore there is also no contribution from the pair of fermions 
at the bottom of the Fermi sea. That is why $\Delta(z)$, contrary 
to the fermion density, reveals a minimum at $z=0$, see
Fig.~\ref{five}.

The analysis of the Bogoliubov modes suggests a simple model of self-localization 
in the case of attractive boson-fermion interactions. Suppose, that in the
vicinity of the localized bosons we may neglect 
the pairing field and the density of all fermions 
except a fermion pair at the bottom of the Fermi sea. 
Then, we obtain the following set of equations  
\bea
(\mu-E_0)v_0&=&\left[-\frac12 \partial^2_z 
-|g_{ff}|v_0^2 - |g_{bf}|N_b\phi^2\right]v_0,
\label{v01}
\\ 
&& \cr
\mu_b\phi&=&\left[-\frac{m_f}{2m_b} \partial^2_z - 2|g_{bf}|v_0^2\right]\phi.
\label{eqsol}
\eea
For 
\be
\frac{m_b}{m_f}=\frac{N_b}{2}+\frac{|g_{ff}|}{2|g_{bf}|},
\label{msol}
\ee
there exists analytical solution 
of Eqs.~(\ref{v01})-(\ref{eqsol}),
\bea\label{soliton}
\phi(z)=v_0(z)=\sqrt{\frac{\alpha}{2}}\;{\rm sech}(\alpha z), 
\eea
with 
\bea
\alpha &=& |g_{bf}|\frac{m_b}{m_f}, \cr 
E_0&=&\mu+\frac{g_{bf}^2m_b^2}{2m_f^2}, \cr
\mu_b&=&-\frac{g_{bf}^2m_b}{2m_f}.
\eea
Such a solution resembles vector solitons. They appear in non-linear optics 
when interactions of several field components are described by a set of
coupled non-linear Schr\"odinger equations \cite{Kivshar}. 
Note that for the 
self-localization of an impurity atom in a large BEC considered in
Ref.~\cite{Sacha2006}, 
the 1D system is described by a parametric soliton with the state of 
the impurity atom given by the hyperbolic secant squared function. 

A comparison of the analytical solutions (\ref{soliton}) with numerical results of 
the full set of equations is shown 
in Figs.~\ref{five}-\ref{six}. The agreement is 
very good and increases with the strength of boson-fermion interactions. 
Indeed, for the strong interaction, due to the Pauli exclusion rule, there 
is negligible probability density 
to find other fermions than the localized pair in the vicinity of $z=0$.
As a consequence, the localized bosons interact almost exclusively with the localized 
fermion pair and the set of Eqs.~(\ref{v01})-(\ref{eqsol}) becomes exact.

Figure~\ref{six}b shows that the Bogoliubov mode $v_1(z)$ forms an 
antisymmetric bound state. In the vicinity of $z=0$ (where the fermion 
density is dominated by $v_0^2$ and the pairing function drops to zero)
this mode should fulfill 
equation similar to Eq.~(\ref{v01}), that is
\be
(\mu-E_1)v_1=\left[-\frac12 \partial^2_z 
-|g_{ff}|v_0^2 - |g_{bf}|N_b\phi^2\right]v_1.
\label{v1}
\ee  
If $\phi$ and $v_0$ are given by Eq.~(\ref{soliton}) the 
antisymmetric solution of Eq.~(\ref{v1}) forms a marginal bound state
\bea
v_1(z)&\sim& \tanh(\alpha z), \\
E_1&=&\mu.
\eea
In the full description of the system, the state govern by the equation (19) 
may become either truely bound or unbound. 
In the considered system, it turns out that the state is pushed towards a true bound state as visible in
Fig. 6b.

When the boson-fermion coupling constant $g_{bf}$ is decreased, we observe the increasing 
discrepancy between analytical and numerical solutions, see Fig.~\ref{seven}. The width of the boson probability density obtained numerically
is significantly greater 
than the corresponding analytical value. This is due to the fact, that in 
the effective potential experienced by the bosons a considerable contribution comes
from other fermions, and not only from the pair at the bottom of the Fermi sea. The density of 
such fermions, contrary to the localized fermion pair, 
possesses a minimum at $z=0$ 
and thus effectively makes the potential well for bosons weaker.
Consequently, bosons occupy a much larger space than can be expected on the basis 
of solutions Eq.~(\ref{soliton}).

\section{Conclusions}
\label{conclusions}

We have considered a small number of bosons immersed in a superfluid 
mixture of fermions in two different spin states. With negligible boson-boson 
interactions, homogeneous densities of the particles become unstable as soon as the 
boson-fermion coupling constant is non-zero. It corresponds to the
phase separation transition. We show that in 3D space for sufficiently
strong repulsive boson-fermion interactions another transition takes place,
i.e. the  self-localization of Bose particles. 
That is, the repulsion between particles creates a local potential well 
for bosons where, if the well is sufficiently large, 
they can localize. The self-localization is present both for the 
superfluid and the normal state of fermions. It modifies properties of the Fermi
sub-system locally without destroying the superfluidity. Low non-zero
temperature affects the pairing function but has little effect on the
self-localization phenomenon.

We do not observe the self-localization for attractive boson-fermion 
interactions in the 3D case. In this context the self-localization requires
sufficiently strong boson-fermion interactions. However, for strong 
attractive interactions no metastable state of the system exists and the
densities of the atoms collapse to Dirac-delta distributions indicating a breakdown 
of the description with the contact interaction potentials.
In the 1D case there is no collapse for attractive boson-fermion
interactions.
The self-localization of bosons is accompanied by localization of a pair of
fermions at the bottom of the Fermi sea. This phenomenon can be described by
a simple model where the self-localization is related to the existence of a
vector soliton solution.

To realize experimentally the self-localization of bosons in a Fermi system,
ultra-cold clouds of bosons and fermions have to be prepared in a laboratory. 
For a sufficiently large boson-fermion coupling constant, that can be achieved 
by means of a Feshbach resonance, the self-localization takes place. Signatures
of the self-localization can be visible in expansion of the atomic clouds
after trapping potentials are turned off. That is, if during the time of flight the
boson-fermion interactions are kept negligibly weak, the initially strongly
localized boson cloud will show much faster expansion than the Fermi cloud due
to release of a large kinetic energy. The simplest experiment would employ 
a Fermi sub-system in a normal phase. In order to observe the self-localization in
a superfluid Fermi mixture a manipulation of a fermion-fermion coupling constant 
is also needed and two Feshbach resonances must be employed, e.g. one resonance
controlled by magnetic field and the other by optical means.

\section*{Acknowledgments}
This work is supported by the Polish Government within research projects
 2009-2012 (KT) and 2008-2011 (KS).


\end{document}